\documentclass[twocolumn,aps,prl,floatfix]{revtex4}
\usepackage[dvips]{graphicx}
\usepackage[english]{babel}
\usepackage{amsmath}
\usepackage{amssymb}

\newcommand{\be}{\begin{equation}}
\newcommand{\ee}{\end{equation}}

\begin{document}

\title{Beyond linear response spectroscopy of ultracold Fermi gases}
\author{J. Kinnunen$^\dagger$ and P. T\"orm\"a$^{\dagger*}$}
\affiliation{$^\dagger$Department of Physics, Nanoscience Center, P.O.Box 35,
  FIN-40014 University of Jyv\"askyl\"a, Finland \\
$^*$Institute for Quantum Optics and Quantum Information of
  the Austrian Academy of Sciences, A-6020 Innsbruck, Austria}

\begin{abstract}
We study RF-spectroscopy of ultracold Fermi gas by going beyond the
linear response in the field-matter interaction. Higher order
perturbation theory allows virtual processes and energy conservation
beyond the single particle level. We formulate an effective higher
order theory which agrees quantitatively with experiments on the
pairing gap, and is consistent with the absence of the mean-field
shift in the spin-flip experiment. 
\end{abstract}

\maketitle

A single atom interacting with a coherent field displays various
coherent phenomena, such as Rabi oscillations, which can be
theoretically described by the exact non-perturbative solution of the
quantum time evolution of the system~\cite{CohenTannoudji}. In many
other contexts, however, it is sufficient to treat the field-matter
interaction within the linear response, i.e.\ to first order in
perturbation theory; this allows one, for instance, to calculate the
response of matter in a complicated many-body state~\cite{Mahan}. In
ultracold Fermi gases of alkali atoms, the interaction between atoms
and fields may be extremely coherent while, at the same time, the
atoms can be strongly interacting or in a non-trivial many-body state
such as superfluid of Cooper pairs. This is likely to lead to new
phenomena in the field-matter interactions. The first indication of
this was the absence of mean-field shifts observed~\cite{Gupta2004} in
the spin-flip experiment using RF fields~[4,3]. 
%\cite{Regal2003}\cite{Gupta2004}.  
The experimental results were in contradiction with predictions given
by linear response. However, they were elegantly
explained~\cite{Gupta2004,Zwierlein2003} as a collective, coherent
rotation of the Bloch vector within a framework that can be applied to
Hartree-type mean-field interactions. It has now become topical to ask
what happens in a similar situation but for a more complicated
many-body state, such as the recently observed Fermi
superfluids~\cite{review}. RF-spectroscopy was used for detecting the
pairing gap in such systems~\cite{Chin2004}. The results were
qualitatively in good agreement~\cite{Kinnunen2004b,Ohashi2004,He2005}
with the lowest order perturbation theory. 

In this letter we combine higher order perturbative approach of the
field-atom interaction with the idea of a collective coherent rotation
of all atoms~\cite{Zwierlein2003} and apply it to the scheme used in 
the experiment~\cite{Chin2004}. The key point is that only total
energy conservation is demanded instead of energy conservation in each
single-atom process. This approach allows the transfer of large
numbers of atoms, as observed in the experiment~\cite{Chin2004} and is
consistent with the absence of mean-field shifts in the spin-flip
experiment~\cite{Zwierlein2003}. It is a step towards high precision
description of the RF-spectroscopy of superfluid Fermi gases, and it
highlights that ultracold Fermi gases may display new phenomena in
field-matter interactions.           

\begin{figure}
  \centering
  \includegraphics[height=3cm]{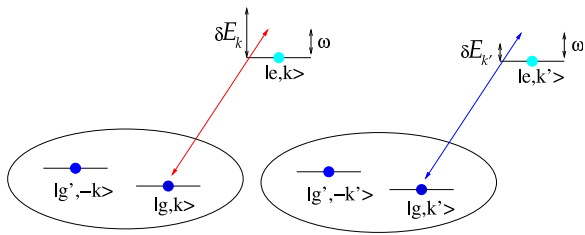}
  \caption{
The RF-field breaks a pair of atoms in internal states $g'$ and $g$ by driving a transition to a
third state. At the single-atom level, the energy cost of creating an excitation in the superfluid, $\delta E_k$,
has to match the RF-field detuning $\omega$, causing a shift of the spectral peak. Higher order perturbation
theory allows processes where energy is conserved at the many particle level, e.g. resonant transfer of the
two atoms above even when they are individually out of resonance.}
  \label{fig:rfscheme}
\end{figure}

We study the RF-spectroscopy of a trapped Fermi gas, as described in
Fig.~\ref{fig:rfscheme}, using perturbative expansion of the
field-atom coupling. The interaction is described in the rotating wave 
approximation, for a spatially constant field with the Rabi frequency
$\Omega$, by 
\be
  H_\mathrm{I} = \Omega \int d{\bf r}\, \left[ e^{i\omega t} \Psi_e^\dagger ({\bf r}) \Psi_g ({\bf r}) + h.c. \right]
\label{eq:fieldatom},
\ee
where $\omega$ is the detuning of the RF-field and the field operators
are expanded in the eigenstates of the harmonic trap potential
$\Psi_\sigma ({\bf r}) = \sum_{nlm} \psi_{nlm} ({\bf r}) c_{nlm
  \sigma}$. 
In the following, the term "momentum conservation" should be
understood, in case of a nonuniform gas, as the conservation of the
trap quantum number. 

\begin{figure}
  \centering
  \includegraphics[height=3cm]{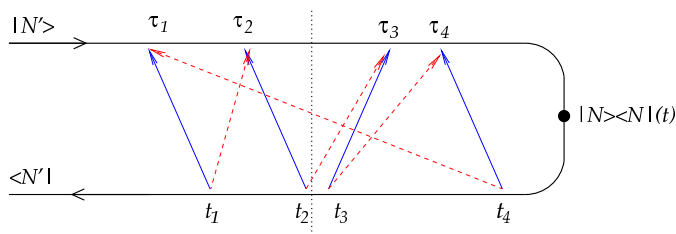}
  \caption{In the sequential transfer (solid arrows) of $N=4$ atoms, the
  (time)order in which the atoms are transferred can be chosen in $N!$
  different ways. 
  An $N$-atom transfer (one example shown by
  dashed arrows) can be done in $N!^2$ different ways. The process is
  called virtual if an auxiliary vertical line (example
  shown as dotted line) cuts two or more arrows~\cite{Schoeller1994}.}
  \label{fig:diagrams}
\end{figure}

The lowest order perturbation theory in $H_\mathrm{I}$ using
the exact trap states gives the transfer probability 
$P_i = \int d{\bf r} \int d{\bf r'} \, P(i,{\bf r}, {\bf r'})$ 
for atom $i = (n_i, l_i, m_i)$. 
Here
\be
\begin{split}
   P(i,{\bf r},{\bf r'}) = &\frac{\Omega^2}{\hslash^2}
   \int_0^t d\tau_1 \int_0^t dt_1 \, e^{i \omega
     (t_1-\tau_1)} \sum_{jj'}
   \Psi_{ijj'} ({\bf r},{\bf r'}) \\ 
&\times \langle c^\dagger_{ig} (\tau_1)
   c_{jg} (t_1) \rangle \langle c_{j'e} (\tau_1)
   c_{j'e}^\dagger (t_1) \rangle,
\label{eq:1order2}
\end{split}
\ee
where the function  $\Psi_{ijj'}
({\bf r},{\bf r'}) = \psi_i
   ({\bf r}) \psi_{j'} ({\bf r}) \psi_j ({\bf r'}) \psi_{j'}
   ({\bf r'})$ describes the overlap of atom wavefunctions.
The expectation value for the number of transferred atoms in the
first-order perturbation theory is obtained by summing Eq.(\ref{eq:1order2})
over all atom indices $i = (n_i,l_i, m_i)$. 
In addition, the integration over the times $\tau_1$ and $t_1$ gives
the energy conservation condition for a single-atom process.
We consider the term of the order $N$ in the perturbation expansion,
(c.f. Fig.~\ref{fig:diagrams})
\begin{widetext}
\be
 \langle \hat N_e (t) \rangle_N = \frac{1}{\hslash^{2N}} \langle  \int_0^t dt_1\, \ldots \int_0^{t_{N-1}} dt_N \, H_I(t_N) \ldots H_I(t_1) N_e(t)
\int_0^t d\tau_N\, \ldots \int_0^{\tau_2} d\tau_1 \, H_I(\tau_N) \ldots H_I(\tau_1) \rangle.
\label{eq:number2}
\ee
\end{widetext}
We take this to be the expectation value for $N$ atoms being
transferred in a single process. Thus, we neglect oscillations of
the atoms back and forth corresponding to terms of higher order in
$H_I$ but with the same final number of transferred atoms. In
principle, these oscillations contribute to the linewidth of
the field-atom interaction. 


For weak excitations (linear response), the single-particle energy
conservation gives the transfer probability $\frac{\Gamma^2}{(\delta E_k - \omega)^2 +
  \Gamma^2}$ for an atom with energy cost $\delta E_k$, where $\omega$
is the field detuning and $\Gamma$ is the linewidth. For strong
single-particle excitations, i.e. Rabi oscillations, this becomes
$\frac{\Omega^2}{(\delta E_k - \omega)^2 + \Omega^2}$, thereby the
field-atom coupling $\Omega$ acts as an effective linewidth. Large
effective linewidth ($\Gamma$ or $\Omega$) allows to transfer a large
number of atoms non-resonantly also in single-particle
processes. However, the reported linewidth in~\cite{Chin2004} is
small, $\sim 100\,\mathrm{Hz}$, which could also be deduced from the
sharpness of the narrowest features of the spectra, giving a limit to
homogenous broadening (other features show inhomegenous broadening due
to pairing and the trap potential). The effective linewidth needed for
the transfer of the observed number of atoms can be estimated, using
the calculated dispersion $\delta E_k$, to be $\sim
10\,\mathrm{kHz}$. Since this is significantly larger than
$100\,\mathrm{Hz}$, considerations beyond single-particle processes
are well motivated.

%The Nth order process may occur via virtual states as shown in
%Fig.~\ref{fig:diagrams}. These states do not need to conserve the
%total energy of the system, as long as the lifetime of the state is
%short as described by the energy uncertainty relation $\Delta E \Delta
%T \geq \hbar$ . This kind of processes are analogous to the
%higher-order phenomenon of cotunnelling in solid state systems, and
%can not be described using the linear response theory.
%
%The relevance of these virtual processes becomes apparent when one
%compares the experimental RF-spectra~\cite{Chin2004} with any theory
%that requires the conservation of single-particle energies.
%The single-particle energy conservation produces a factor
%$\frac{\Gamma^2}{(\delta E_k-\omega)^2 + \Gamma^2}$ for the transfer
%probability of an atom with the energy cost $\delta E_k$ and for field
%detuning $\omega$. For Rabi
%oscillations, the effective linewidth $\Gamma$ is given by the
%field-atom coupling $\Omega$. Increasing the field intensity does
%increase the number of transferred atoms at the cost of increasing the
%linewidth.
%
%On the other hand, the excitation
%gap $\Delta$ of the paired gas broadens the energy spectrum of the
%atoms. In order to obtain transfer rates shown in the
%experiments~\cite{Chin2004} for the peak shifted from the zero
%detuning, one would require linewidths of the order $10^4\,Hz$ while
%the reported linewidths, and the effective linewidths that could be
%deduced from the sharpness of the ``free atom'' peak at the zero
%detuning, were of the order $10^2\,Hz$. 

Assume an effective pulse length $T_\mathrm{P}$ (either the true pulse
length or the coherence time of the interaction). If the number of
transferred atoms $N$ is large, the transfer processes occur at
roughly even intervals. This means that each intermediate virtual state lasts
for roughly the same time $t_\mathrm{inter} =
\frac{T_\mathrm{P}}{N-1}$. For pulse lengths of $T_\mathrm{P} = 1\,s$, 
and the number of transferred atoms of the order $N = 10^5$ the
corresponding energy uncertainty for any single virtual state is
$\Delta E/\hbar \geq \frac{1}{t_\mathrm{inter}} = 10^5\,Hz$ (repeated
transfer of smaller $N$ within coherence-limited $T_\mathrm{P}$ gives
the same result). Since the typical Fermi energies are of the order of
$10\,kHz$, this allows any atom to be transferred regardless of
its energy cost. However, the total Nth order process should conserve
energy, meaning that the sum of $\delta E_k - \omega$ of all
transferred atoms should be zero.  

The effective pulse length $T_\mathrm{P}$ enters the picture only through the
uncertainty relation, by introducing an energy cutoff
$\delta_\mathrm{c}$ for the maximum energy cost of an atom
transferred through an intermediate (virtual) state. 
This allows to simplify Eq.~(\ref{eq:number2}) by fixing the cutoff
$\delta_\mathrm{c}$. Thereby the actual duration of the virtual state
$t_\mathrm{inter}$ becomes irrelevant for $t_\mathrm{inter} \leq
\frac{\hbar}{\delta_\mathrm{c}}$ and we can choose $t_\mathrm{inter}
\rightarrow 0$. 
Considering the RF-spectroscopy in the sense of a local density
approximation, calculating the spectra for each position ${\bf r}$
separately and then integrating over ${\bf r}$, we also demand that
the atoms may not exchange energy in the Nth order process over a
distance. The probability of finding the atoms at the same position is
given by the overlap of the atom wavefunctions. These approximations
can be written as 
\be
 \langle \hat N_e (t) \rangle_N \approx \frac{t^{2N-2}}{\hslash^{2N}} \int_0^t d\tau_1 \int_0^t dt_1 \langle
H_I^N(\tau_1) N_e(t) H_I^N(t_1) \rangle
\label{eq:Norder0}
\ee
where
\be
  H_I^N (t) =  C \int d{\bf r}\, \left[
    \Omega e^{i\omega t} \Psi_e^\dagger ({\bf r},t) \Psi_g ({\bf r},t) +
    h.c.\right]^N,
\label{eq:Nfieldatom}
\ee
and the normalisation constant $C = V^{N-1}$. The integration over the
position ${\bf r}$ gives the total momentum conservation. 
Having only two time variables in Eq.(\ref{eq:Norder0}) is related to
the simultaneous transfer approximation ($t_\mathrm{inter} \rightarrow
0$), and the single 
position variable in Eq.(\ref{eq:Nfieldatom}) corresponds to the
requirement of the wavefunction overlap. 
For fermions, the operator $H_I^N$ as expressed above vanishes for $N>1$ because
of the Pauli exclusion principle. However, restricting oneself to a
subset where each one-atom transfer preserves momentum separately, one
obtains a nonvanishing contribution even for fermions. That is, we replace
\be
  \Psi_e^\dagger ({\bf r},t) \Psi_g ({\bf r},t) \rightarrow \sum_{nlm} 
\left| \psi_{nlm} ({\bf r}) \right|^2 c_{nlme}^\dagger (t)
c_{nlmg} (t).
\ee
The process described here is therefore only possible in presence of
momentum conservation in the field-matter interaction, as is indeed
the case in the considered system.

Our Nth order theory is now simply a linear response theory for the
defined coherent N-atom coupling operator $H_I^N(t)$. One obtains using
the Wick's theorem (when no atoms in state $e$ initially)
\be
  \langle \hat N_e(t,{\bf r}, {\bf r'}) \rangle_N = C' 
  \int_0^t d\tau_1 \int_0^t dt_1 \, e^{iN\omega (t_1 - \tau_1)}
\sum_{\{i\}} P_{\{ i \}}
\ee
where the prefactor $C' =
\frac{NC^2t^{2N-2}\Omega^{2N}}{\hslash^{2N}}$, the summation is over
all $N$-atom sets $\{ i \}$ and the probability that this set of atoms
$\{ i \}$ is transferred is
\be
 P_{\{ i\} } = \sum_{\overline{\{j\}}}
\prod_{i,j\leq N} \Psi_{ij} ({\bf r},{\bf r'}) \langle c_{ig}^\dagger (\tau_1) c_{jg} (t_1) \rangle \langle c_{ie}
(\tau_1) c_{ie}^\dagger (t_1)  \rangle.
\ee
Here $\overline{\{j\}}$ denotes all permutations of the set $\{ i\}$.
Note that even though each one-atom process
preserves momentum, the Nth order process still contains off-diagonal
Green's functions $\langle c_{nlmg}^\dagger c_{n'l'm'g}
\rangle$.

We approximate the sum over all permutations $\overline{\{ j \}}$ by $N$
independent sums, thus replacing the $N!$ terms by $N^N$ terms. This
approximation contains all the correct terms but it also introduces
additional terms. 
However, the diagonal Green's functions dominate over
the off-diagonal ones, and these additional terms give only
a small contribution. In addition, the product over the
functions $\Psi_{ij}$ gives cancellations of the added
terms upon ${\bf r}$-integration since, for these terms, the atom wavefunctions do not necessarily appear in the square form $|\Psi_{ij}|^2$. 
One obtains the following form for the transfer probability of set $\{ i \}$
\be
  P_{\{ i\} } = \prod_{i\leq N} \sum_j \Psi_{ij} ({\bf r},{\bf r'}) \langle c_{ig}^\dagger (\tau_1) c_{jg} (t_1) 
\rangle \langle c_{ie} (\tau_1) c_{ie}^\dagger
(t_1) \rangle.
\label{eq:Norder1}
\ee
This probability can be formally written at $T=0$ as
\be
  P_{\{ i\}} = \prod_{i \leq N} \sum_j \sum_m e^{i\Delta
    E(i,j,m)(\tau_1-t_1)} F(i,j,m,{\bf r},{\bf r'}),
\label{eq:Norder2}
\ee
where $\Delta E(i,j,m)$ is the energy difference of the single-atom
states and $F(i,j,m)$ is a product of Fermi functions and Bogoliubov
coefficients giving the occupation and transfer probabilities,
irrespective of the energy conservation. The internal
variable $m$, describing the quasiparticle states, comes from the
Bogoliubov-deGennes equations. Comparison of Eq.(\ref{eq:Norder1}) to Eq.(\ref{eq:1order2})
shows that the first-order term contains the same spectral weight factors,
if we choose $j = j'$ in Eq.~(\ref{eq:1order2}). The sum in Eq.~(\ref{eq:1order2})
becomes 
\be
  \sum_j \sum_m e^{i\Delta
    E(i,j,m)(\tau_1-t_1)} F(i,j,m,{\bf r},{\bf r'}).
\label{eq:Norder3}
\ee
Integration over the times $\tau_1$ and $t_1$ gives the single atom
energy conservation. In (\ref{eq:Norder2}), the exponential functions
in the product over $i$ can be combined 
and upon integration over times $\tau_1$ and $t_1$ in
Eq.(\ref{eq:Norder0}), yield the total energy conservation. This acts
as a boundary condition for the choice of the transferable atoms
$\{ (n_i, l_i, m_i) \}$ and hence for the actual number of transferred 
atoms $N$. The expectation number of transferred atoms is now the
average of all combinations $\{ \mathcal{A} \}$ that satisfy the total
energy conservation. The spectral weights, or the probabilities,
of different combinations are determined by the product of the terms
$F(i,j,m,{\bf r},{\bf r'})$ which also appear in the first order spectra in
Eq.~(\ref{eq:Norder3}). We approximate the expectation value by keeping
in $\{ \mathcal{A} \}$ only the combination that transfers the highest number
of atoms. We have tested this approximation with $40$ atoms and found
good agreement between the approximative scheme and the full
combinatorial treatment. In calculating the total energy conservation,
the energy cutoff $\delta_\mathrm{c}$ for the maximum energy cost of
an transferrable atom acts as an additional constraint for the
possible combinations.

We now use the first order spectrum to calculate the number of
transferred atoms by the following algorithm.
%We now calculate the number of transferred atoms by the following algorithm.
%It uses the first order spectrum to pick the largest possible set of
%atoms that can be transferred while still satisfying the total energy conservation.

\begin{itemize}
\item Consider a fixed field detuning $\omega$. Set $\delta_{min} = -\infty$, $\delta_{max} = \infty$. Let
      $N_1(\delta)$ be the expectation number of atoms that one can transfer using
      the first-order perturbation theory and by giving energy $\delta$. 
\item Integrate $\int_{\delta_{min}}^{\delta^{max}} d\delta\,
      N_1(\delta) (\delta-\omega) =: \Delta E(\delta_{min},\delta_{max},\omega)$.
      This gives the total energy change if all the atoms with energy change of $\delta \in \left[ \delta_{min}, \delta_{max}\right]$ are
      transferred.
\item If $\Delta E(\delta_{min},\delta_{max},\omega)$ is positive, decrease $\delta_{max}$ or if it is negative, increase $\delta_{min}$
      and go back to the previous step. Otherwise, proceed to the next step.
\item The maximum number of atoms that one can transfer is $N_N(\omega) := \int_{\delta_{min}}^{\delta_{max}} d\delta\,N_1(\delta)$.
\end{itemize}
One can map the whole spectrum by using the algorithm above
for several detunings $\omega$. For a nonuniform gas, the
process should be repeated for all positions ${\bf r}$ and ${\bf r'}$ 
and integrated to give the total number of transferred atoms.

For a noninteracting gas, the first-order perturbation theory and the
current higher order theory give identical results
since the conservation of the total energy equals the conservation of energy of a single one-atom
process. For atoms interacting only via the Hartree field, 
the two theories differ by the position of the spectral peak.
As an example we consider the spin-flip experiment of Ref.~\cite{Zwierlein2003},
with $N_g$ ($N_{g'}$) atoms in the $g$-state
($g'$-state) and mutual interaction energy $V_{gg'}$.
In the first-order perturbation theory, the RF-field needs
to give the energy corresponding to a single atom
excitation, i.e.\ $\Delta E
= V_{gg'} (N_g-N_{g'})$. 
%For large
%interaction strengths this energy shift can be large. One should also
%notice that the first-order perturbation theory does not predict any
%kind of atom transfer after the equilibrium state of equal atom
%numbers in both atom states is achieved. 
The higher order theory
requires only the energy corresponding to the complete $N$-atom
process. Since the mutual interaction energy of the two fermion
species $g$ and $g'$ is unchanged if the spins of all
atoms are flipped, the total energy change of the whole process is
zero \cite{selitys}. 
This is consistent with the absence of the mean-field shift observed in the
spin-flip experiment in Ref.~\cite{Zwierlein2003}.

To describe the superfluid Fermi gas in a harmonic trap, we apply the
mean-field resonance superfluidity approach used in
Ref.~\cite{Ohashi2004} but restricting oneself to the one-channel
model in order to describe broad Feshbach-resonances~\cite{Bruun2001,Ohashi2005}.
We calculate self-consistently the position dependent excitation gap, density 
distribution and the chemical potential for 2670 atoms using
background interaction strength $U = 2.3\,\hslash
\omega_0 r_\mathrm{osc}^3$ and cutoff $E_c = 161.5\,\hslash \omega_0$,
where $\hslash \omega_0$, $r_\mathrm{osc}$ are the oscillator energy
and the length. The position dependent excitation gap is
$0.83\,E_\mathrm{F}$ at the center of the trap corresponding roughly
to Li$^6$ atoms with peak density of $10^{13}\, 1/\mathrm{cm}^3$ at
the magnetic field of $B \approx 834\,\mathrm{G}$, as obtained from
the renormalised theory for a uniform gas~\cite{Kinnunen2004b}. 
The small number of atoms used above gives too narrow density and
gap profiles, and one expects a larger excitation gap close
to the edges of the trap for numbers of atoms used in the experiments.
On the other hand, for pulses longer than the coherence time of the
atoms, the system can relax and lower the excitation gap during the
pulse (not considered here). Thus, one expects the two approximations
to partly cancel each other.

\begin{figure}
  \centering
  \includegraphics[height=5cm]{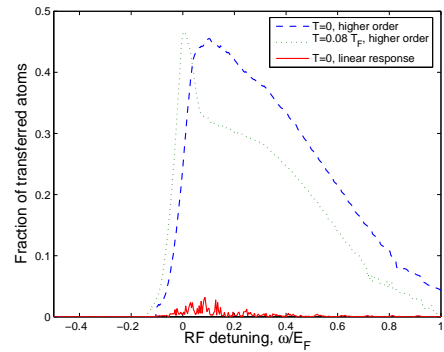}
  \caption{RF-spectra in the higher order theory using parameters
  given in the text. At finite temperatures, an additional peak
  appears at the zero detuning. For comparison, the spectrum from the
  linear response theory is shown in solid line.}
  \label{fig:norder1}
\end{figure}

Fig.~\ref{fig:norder1} shows the RF-spectra at two temperatures $T =
0.0\,T_\mathrm{F}$ and $0.08\,T_\mathrm{F}$ using a
$\pi/2$-pulse. 
The peak is shifted due to the pairing gap and, most notably, another
peak at the zero detuning exists at a finite temperature, originating
from contributions from the edges of the
trap~\cite{Kinnunen2004b}. This behaviour and the locations of the two
peaks at finite temperatures are in good quantitative agreement with the
experiment~\cite{Chin2004}.
%%This temperature behaviour as well as the
%%shape of the finite temperature curve are in good qualitative
%%agreement with the experiment \cite{Chin2004}.
Note that, in principle, first-order perturbation theory gives a
maximum transfer ratio of only $\sim 5\%$ of the atoms, because the
single particle excitation energies $\delta E_k$ have a strong
momentum ($k$) dependence and therefore only a small fraction of the
atoms can match the energy of a narrow linewidth field. The higher
order theory is not limited in this way and allows to describe the
experiments with large numbers of transferred atoms, such as $50\%$
in~\cite{Chin2004}; c.f.\ Fig.~\ref{fig:norder1}.

In conclusion, 
the connection between the first-order and higher
order terms in perturbation theory was used to define
an effective higher order theory for the RF-spectroscopy of Fermi gases. 
It was shown to be consistent with the temperature dependence of the
pairing gap spectra \cite{Chin2004} and the coherence phenomenon
observed in Ref.~\cite{Zwierlein2003}. This approach could be applied
also to other types of field-matter interactions and spectroscopies in
ultracold gases. It should be useful whenever the single particle
excitation spectrum is non-trivial, e.g.\ BCS-type states or exotic
states in optical lattices, and the process is coherent. Due to the
special nature of field-matter interactions discussed here, the
understanding of phenomena such as Josephson oscillations in Fermi
gases --- when the effective tunneling is realized by fields
\cite{Sorin} --- will be non-trivial and may show phenomena not
existent in the corresponding solid state system. 

{\it Acknowledgements} P.T.\ thanks the Aspen Center for Physics. 
This project was supported by Academy of Finland and EUROHORCs (EURYI award,
Academy project numbers 106299, 205470), and the QUPRODIS project of
EU.


\begin{thebibliography}{4}
\bibitem{selitys}
It is always this energy change that appears in the exponential terms in Eq.(\ref{eq:Norder2}). How many spins are actually flipped depends on the %%@
factor in front, $\Omega t$, i.e. whether $\pi/2$ pulse or shorter is applied, but the energy term contains the final energy corresponding to all %%@
spins flipped.
\end{thebibliography}
\end{document}